\pdfoutput=1

\documentclass[10pt,letterpaper]{article}
\usepackage[top=0.85in,left=2.75in,footskip=0.75in]{geometry}

\usepackage{changepage}

\usepackage[utf8]{inputenc}

\usepackage{textcomp,marvosym}

\usepackage{fixltx2e}

\usepackage{amsmath,amssymb}

\usepackage{cite}

\usepackage{nameref,hyperref}

\usepackage[right]{lineno}

\usepackage{microtype}
\DisableLigatures[f]{encoding = *, family = * }

\usepackage{rotating}

\usepackage{bm}

\raggedright
\usepackage[aboveskip=1pt,labelfont=bf,labelsep=period,justification=raggedright,singlelinecheck=off]{caption}

\makeatletter
\renewcommand{\@biblabel}[1]{\quad#1.}
\makeatother

\date{}

\usepackage{lastpage,fancyhdr,graphicx}
\fancyheadoffset[L]{2.25in}
\fancyfootoffset[L]{2.25in}

\begin{document}
\vspace*{0.35in}

\begin{flushleft}
{\Large
\textbf\newline{New Routes to Phylogeography}
}
\newline
\\
Nicola De Maio\textsuperscript{1,2},
Chieh-Hsi Wu\textsuperscript{2},
Kathleen M O'Reilly\textsuperscript{3},
Daniel Wilson\textsuperscript{1,2,4,*}
\\
\bf{1} Institute for Emerging Infections, Oxford Martin School, Oxford, United Kingdom
\\
\bf{2} Nuffield Department of Medicine, University of Oxford, Oxford, United Kingdom
\\
\bf{3} MRC Centre for Outbreak Analysis and Modelling, School of Public Health, Faculty of Medicine, Imperial College London, London, United Kingdom
\\
\bf{4} Wellcome Trust Centre for Human Genetics, University of Oxford, Oxford, United Kingdom
\\

%
%





* E-mail: daniel.wilson@ndm.ox.ac.uk 
\end{flushleft}
\section*{Abstract}

Phylogeographic methods aim to infer migration
trends and the history of sampled lineages from genetic data. Applications
of phylogeography are broad, and in the context of pathogens include
the reconstruction of transmission histories and the origin and emergence
of outbreaks. Phylogeographic inference based on bottom-up population
genetics models is computationally expensive, and as a result faster
alternatives based on the evolution of discrete traits have become
popular.
In this paper, we show that inference
of migration rates and root locations based on discrete trait models
is extremely unreliable and sensitive to biased sampling. To address
this problem, we introduce BASTA (BAyesian STructured coalescent Approximation),
a new approach implemented in BEAST2 that combines the accuracy of
methods based on the structured coalescent with the computational
efficiency required to handle more than just few populations.
We illustrate the potentially severe implications of
poor model choice for phylogeographic analyses by investigating the
zoonotic transmission of Ebola virus. Whereas the structured coalescent
analysis correctly infers that successive human Ebola outbreaks have
been seeded by a large unsampled non-human reservoir population, the
discrete trait analysis implausibly concludes that undetected human-to-human
transmission has allowed the virus to persist over the past four decades.
As genomics takes on an increasingly prominent role informing the
control and prevention of infectious diseases, it will be vital that
phylogeographic inference provides robust insights into transmission
history.

\section*{Author Summary}

When studying infectious diseases it is often important to
understand how germs spread from location-to-location, person-to-person,
or even one part of the body to another. Using phylogeographic methods,
it is possible to recover the history of spread of pathogens (or
other organisms) by studying their genetic material. Here we compare
different phylogeographic methods based
on principled population models and fast alternatives. We found that different approaches can give diametrically opposed results, and
we offer a concrete example in the context of the ongoing Ebola outbreak
in West Africa. We found that the most popular phylogeographic method
often produces completely inaccurate conclusions. One of the reasons
of this popularity has been its computational speed, which has
allowed users to analyse large genetic datasets with complex models.
More accurate approaches have until now been considerably slower,
and therefore we propose a new method called BASTA that achieves
good accuracy in a reasonable time. We are relying more and more on genetic sequencing to learn about the origin
and spread of infections, and as this role continues to grow, it will
be essential to use accurate phylogeographic methods when designing
policies to prevent or kerb the spread of disease.


\section*{Introduction}
Phylogeographic methods aim to infer many aspects of population
evolution from genetic data. The phylogeography term often encompasses
methods to infer changes in population sizes (phylodynamics) and population
divergence events (see~\cite{BLOOETAL10}). In this work, we focus
on the inference of migration between distinct subpopulations (such
as in the structured coalescent, see~\cite{NORD97}). For many years,
nested clade phylogeographic analysis (NCPA, see e.g.~\cite{TEMPETAL87,TEMP93})
was the leading method to test isolation and migration (reviewed in~\cite{TEMP10,BLOOETAL10}).
More recently, model-based inference for phylogeography has flourished
and has replaced NCPA as the new standard approach (reviewed in~\cite{HEYMACH03,BEAUETAL10}).

Phylogeographic model-based approaches for migration have predominantly been used to study the spread of pathogens through geographic locations and to infer their original source~\cite{LEMEETAL09,LEMEETAL10,ALLIETAL12,FARIETAL12}, but is also commonly used to study the migration history of animals~\cite{CAMPETAL10,BRANETAL11,EDWAETAL11}, plants~\cite{DRUMETAL12b,LIUETAL12}, and even languages~\cite{BOUCETAL12}.
Phylogeographic methods can also be useful to address a wider range of questions in epidemiology, for example when studying  transmission of pathogens between body compartments in the same host~\cite{CHAIETAL14}, between individual hosts~\cite{DIDEETAL12}, between host social groups~\cite{GRADETAL14}, or between host species~\cite{SPOOETAL13}.

The first class of model-based approaches that we consider
are likelihood-based methods implementing the structured coalescent~\cite{BEERFELS99,BEERFELS01,EWINETAL04,BEER06,VAUGETAL14}.
These approaches use the structured coalescent without approximation
to infer migration rates and effective population sizes. These methods
are not practical in scenarios with large numbers of populations and
migration events due to their computational demand. In fact, they
not only explore the space of parameters of primary interest (such
as migration rates, population sizes, and phylogeny) but also the
space of all possible migration histories, vastly increasing the computational
complexity.

Recently, an alternative phylogeographic approach has been
proposed, which models the evolution of locations as a discrete trait,
in the same way as evolution of genetic loci is usually modeled~\cite{LEMEETAL09,LEMEETAL10,EDWAETAL11}.
Since migration is modelled similarly to genetic mutations, this model
is referred to as ``Mugration'' by~\cite{KUHNETAL11, BOUCETAL14}, or more commonly as ``discrete trait analysis'' (DTA in the following). This approximation
has become very popular (see e.g.~\cite{BLOOETAL10}) thanks also
to its computational efficiency and user-friendly software. Yet, this model is based on a set of approximations that profoundly diverge from
classical models of migration in population genetics (see e.g.~\cite{NOTO90,HERB94,WILK98}).
While methods based on the structured coalescent, which accounts for the effects of migration on the genealogy, are preferable over DTA, the latter is often chosen due to the computational demands of current implementations of the structured coalescent.
DTA is also commonly used to describe the evolution of discrete phenotypes.
In such cases, DTA may be appropriate~\cite{CUNNETAL98,PAGE99}.
However, this requires some assumptions that are usually not met when the studied trait is a geographic location, including 
that trait frequencies in the global population can drift and reach loss or fixation and
that trait sampling is random, i.e., that the number of samples carrying each trait is not pre-determined, but is obtained by sampling randomly from a single population containing all traits.

There is a scarcity of studies in the scientific literature
assessing the accuracy of DTA and comparing different phylogeographic
approaches, but concerns have been raised because, among other other
issues, DTA is thought to be sensitive to the sampling locations
chosen~\cite{FARIETAL12,LEMEETAL14}. Also, because DTA conceptually separates the coalescent
and the migration process, it might lead to non-optimal
use of information. Here we demonstrate that these concerns are well
founded, in that DTA suffers from various biases and statistical inefficiency
despite its efficient approximation.

To address this problem we introduce a new model-based approach that uses a close approximation to the structured coalescent (similar in spirit to~\cite{VOLZ12,  RASMETAL14}).
The idea behind this approximation is to efficiently integrate over the parameter space of all possible migration histories, therefore reducing the computational effort needed to explore the space of parameters of interest.
We implement this approach in the Bayesian phylogenetic package BEAST2~\cite{BOUCETAL14}, and call it BASTA (BAyesian STructured coalescent Approximation).
We compare its performance to DTA and MultiTypeTree (MTT, a recent Bayesian structured coalescent software, see~\cite{VAUGETAL14}) using simulations based on the structured coalescent.

To demonstrate the importance of model choice, we analysed genomic data from previous and ongoing Ebola outbreaks~\cite{GIREETAL14} using different phylogeographic approaches, and studied the contribution of zoonotic events to Ebola contagion.
We show that, based both on simulations and real data analyses, DTA and structured coalescent methods can lead to different conclusions, and in particular DTA is often inaccurate.

\section*{Materials and Methods}

\subsection*{The Structured Coalescent}
Using the notation of~\cite{VAUGETAL14}, we introduce the structured coalescent model in a Bayesian inference setting.
Data from a given set of samples $I$ consist of three elements: $S=\{ s_i | i \in I \}$ the aligned genetic sequences, $t_I=\{ t_i | i \in I \}$ the sampling dates, and $L=\{ l_i | i \in I \}$ the sampling locations.
We are interested in estimating parameters describing the migration and coalescent processes: $\bm{m}$ the matrix of migration rates between demes and $\bm{\theta}$ the vector of effective population sizes for the considered demes.
Lastly, the model has additional ``nuisance'' parameters: $T$ the coalescent tree, $\bm{\mu}$ the substitution rate matrix, and $M$ the migration history of lineages in the tree.
We want to estimate the posterior distribution of the parameters of interest using Bayes' theorem:

\begin{equation}
P(T,M,\bm{\mu},\bm{m},\bm{\theta}|S,t_I,L)\propto P(S|T,t_I,\bm{\mu}) P(T,M|t_I,L,\bm{m},\bm{\theta}) P(\bm{\mu},\bm{m},\bm{\theta}).
\label{StCoal}
\end{equation}

The first term $P(S|T,t_I,\bm{\mu})$ can be computed using Felsenstein's pruning algorithm~\cite{FELS81}. 
The second term represents the probability of the migration and coalescent history under the structured coalescent model given the population genetic parameters (the migration rates $\bm{m}$ and population sizes $\bm{\theta}$).
The last term, $P(\bm{\mu},\bm{m},\bm{\theta})$ represents the prior distribution of the parameters, and may be factored to $P(\bm{\mu})P(\bm{m})P(\bm{\theta})$ assuming independence of $\bm{\mu}$, $\bm{m}$, and $\bm{\theta}$.

We now briefly describe how $P(T,M|t_I,L,\bm{m},\bm{\theta})$ is calculated within the structured coalescent.
We assume that we have a set of unique demes $D$, where each deme $d\in D$ has an effective population size $\theta_d$.
We further assume that $m_{dd'}$ is the backward-in-time instantaneous migration rate from population $d$ to population $d'$.
We consider the sequence of time intervals $I_i, i \in 1 \ldots B$ of length $\tau_i$ between subsequent events (coalescent, sampling, or migration), starting from the most recent sample and going back to the root of the phylogeny.
The contribution to the likelihood of interval $i$ is, assuming haploidy:

\begin{equation}
L_i = \exp \left[-\tau_i \sum_{d \in D}\left( \binom{k_{i,d}}{2}\dfrac{1}{\theta_d} + k_{i,d}\sum_{d'\in D,d'\neq d} m_{dd'} \right) \right] E_i 
\label{like1}
\end{equation}
where $k_{i,d}$ is the number of lineages in deme $d$ in interval $i$, and $E_i$ is the contribution of the event that ends interval $i$:

\begin{equation}
E_i = \left\{  \begin{array}{lcl} 1 & \mbox{if it is a sampling event} \\ 
m_{dd'} & \mbox{if it is a migration event from $d$ to $d'$} \\ 
\dfrac{1}{\theta_d} & \mbox{if it is a coalescence event in deme $d$.}  \end{array}  \right.
\end{equation}
The full likelihood is obtained by taking the product of Eq.~\eqref{like1} over all intervals.\\

In the structured coalescent migration events affecting lineages in the tree are explicitly parameterized and estimated (Fig.~\ref{trees}). 
The next two methods presented, DTA and BASTA, avoid this, therefore reducing statistical complexity and computational demand.

\begin{figure}[h]
\caption{{\bf Graphical representation of phylogeographic models.}
In this study we consider three phylogeographic methods: the structured coalescent, DTA, and BASTA.
This figure shows some of the differences in these models, in particular in the modelled events and time intervals.
Coloured dots show different populations (one red population and one green) for both sampled and internal phylogenetic nodes.
a) In the structured coalescent eight events are considered, delimiting seven time intervals of lengths $\tau_1\ldots \tau_7$.
Three of these events are sampling events (denoted by the grey horizontal lines), one is a migration event (represented by an arrow between two coloured dots), and four are coalescence events.
b) In DTA, migration events are not explicitly parameterised, so we have a total of seven sampling or coalescence events, delimiting six time intervals of lengths $\tau_1\ldots \tau_6$. 
While in the figure locations for internal nodes are depicted, the method effectively integrates over all possible ancestral locations at each MCMC step.
c) As in DTA, BASTA does not consider migration events, and therefore has seven events and six time intervals.
Yet, each of these intervals is split exactly in half (blue horizontal dotted lines), and the two halves are considered separately.
Again, as in DTA, at each MCMC step BASTA integrates over all possible internal nodes locations.
}
\label{trees}
\end{figure}

\subsection*{Discrete Trait Analysis}

Here we give a short introduction to the Bayesian implementation of the discrete trait analysis~\cite{LEMEETAL09,EDWAETAL11}, (DTA).
Similarly to the structured coalescent, the focus is on the inference of migration parameters, but there are some noticeable differences:
all demes have the same effective population size; the effects of migration on the coalescent process are ignored, and a standard coalescent prior is used for the coalescent tree.
As mentioned above, migration is modelled as if it were genetic mutation.
In summary, instead of Eq.\eqref{StCoal}, the DTA model adopts the following approximation:

\begin{equation}
P(T,\bm{\mu},\bm{m},\bm{\theta}|S,t_I,L)\approx P(S|T,t_I,\bm{\mu}) P(L|T,t_I,\bm{m}) P(T|t_I,\bm{\theta}) P(\bm{\mu},\bm{m},\bm{\theta})
\label{Mugration}
\end{equation}
where again $P(\bm{\mu},\bm{m},\bm{\theta})$ may be factored to $P(\bm{\mu})P(\bm{m})P(\bm{\theta})$, both likelihoods $P(S|T,t_I,\bm{\mu})$ and $P(L|T,t_I,\bm{m})$ are calculated with the pruning algorithm, and $P(T|t_I,\bm{\theta})$ is simply a neutral coalescent prior for an unstructured population.
The consequences of this approximation have not been thoroughly explored in the literature, despite the popularity of DTA.
One concern is that the effects of migration rates on the shape of the phylogeny are ignored.
For example, when migration rates are very low, long branches close to the root are expected.
This type of information is ignored by DTA, and this could lead to reduced accuracy.

Furthermore, sampling locations $L$ are treated as data (with likelihood $P(L|T,t_I,\bm{m})$), similarly to genetic alignment, while in the structured coalescent they are conditioned upon as auxiliary variables in the tree prior $P(T,M|t_I,L,\bm{m},\bm{\theta})$, in a similar manner to sample size.
Sampling locations are usually arbitrarily chosen by the investigator, and are not a consequence of the migration process, as assumed in DTA.
This could upwardly bias estimated migration rates towards over-sampled populations, and we explore this eventuality using simulations.
An effect of sampling intensity over migration rates and ancestral locations estimated using DTA has already been shortly reported elsewhere~\cite{FARIETAL12,LEMEETAL14}, but has never been explored in detail.

DTA is also commonly used to describe the evolution of discrete phenotypes (in this sense, it is more appropriately referred to as ``discrete trait analysis'').
This model has not to be considered an approximation in such cases, and we do not refer to those cases here, but we only consider when DTA are used to model migration.

\subsection*{BASTA}

We pursue an approximation to the Bayesian structured coalescent that is both accurate and computationally efficient.
Similarly to Eq \ref{StCoal} we define:

\begin{equation}
P(T,\bm{\mu},\bm{m},\bm{\theta}|S,t_I,L)\propto P(S|T,t_I,\bm{\mu}) P(T|t_I,L,\bm{m},\bm{\theta}) P(\bm{\mu},\bm{m},\bm{\theta}).
\label{BASTA}
\end{equation}
where $P(S|T,t_I,\bm{\mu})$ and $P(\bm{\mu},\bm{m},\bm{\theta})$ are calculated as in the structured coalescent.
However, we marginalise over $M$, and instead of calculating the joint likelihood of the coalescent and migration history $P(T,M|t_I,L,\bm{m},\bm{\theta})$, we approximate the likelihood of the coalescent history alone $P(T|t_I,L,\bm{m},\bm{\theta})$ by integrating over all possible migration histories $M$.
This reduces the parameter space at the cost of introducing an approximation similar to~\cite{VOLZ12, RASMETAL14}.

As in the structured coalescent, we consider a sequence of events backward in time, but now they can only be sampling or coalescence events (Fig.~\ref{trees}).
These events define a sequence of time intervals $I_i, i \in 1 \ldots B$ of lengths $\tau_i$ between successive events.
 We describe here the contribution to the likelihood of any single interval $I_{i}=[\alpha_{i-1} , \alpha_i]$ (with $|\alpha_{i-1}-\alpha_i|=\tau_i$, and $\alpha_{i}$ being the older event time of $I_i$ and $\alpha_{i-1}$ the more recent one). 
 The total likelihood is then obtained by multiplying the contribution from each interval.
The likelihood of the interval $I_i$, that we want to approximate, is:
\begin{equation}
L_{i} = \exp \left[-\int_{\alpha_{i-1}}^{\alpha_i} \sum_{d \in D} \sum_{l \in \Lambda} \sum_{l' \in \Lambda, l'\neq l} P(l\in d, l' \in d | t) \dfrac{1}{\theta_d} dt \right] E_i
\label{trueRate}
\end{equation}
where $\Lambda$ is the set of all extant lineages during $I_i$, and $P(l\in d, l' \in d | t)$ is the probability that lineages $l$ and $l'$ are in these same deme $d$ at time $t$ (we are implicitly conditioning on the more recent coalescent history).
$E_i$ is the contribution of the coalescent or sampling event (defined later).
Eq.~\eqref{trueRate} is different from Eq.~\eqref{like1} in that we are considering intervals delimited only by sampling or coalescence events, and not by migration events.
In fact, Eq.~\eqref{trueRate} integrates over all possible migration histories.

The first approximation we do is to replace $P(l\in d, l' \in d | t)$ by $P(l \in d |t)P(l' \in d |t)$, that is, modeling the migration of lineages as independent from one another.
We will call $P_{l,t}$ the vector whose element $d$ is $P_{l,t,d}:=P(l \in d |t)$.
Our further approximation is to split $I_i$ into two sub-intervals of equal length $I_{i1}=[\alpha_{i-1} , (\alpha_i + \alpha_{i-1})/2]$ and $I_{i2}=[(\alpha_i + \alpha_{i-1})/2 , \alpha_i]$, such that $P_{l,t}$ is approximated as $P_{l,\alpha_{i-1}}$ for all $t$ in $I_{i1}$ and $P_{l,\alpha_i}$ for all $t$ in $I_{i2}$.
Then the approximated likelihood contribution of $I_{i1}$ becomes:
\begin{equation}
\tilde L_{i1} = \exp \left[-\dfrac{\tau_i}{2} \sum_{d \in D} \sum_{l \in \Lambda} \sum_{l' \in \Lambda, l'\neq l} P_{l,\alpha_{i-1},d} P_{l',\alpha_{i-1},d} \dfrac{1}{\theta_d} \right] 
\label{rate}
\end{equation}
and similarly is defined $\tilde L_{i2}$.

$P_{l,t}$ is a row vector of dimension $D$, and if $l$ is sampled at time $t$, $P_{l,t}$ is an indicator vector representing the deme of sampling.
Given $P_{l,\alpha_{i-1}}$ we can calculate $P_{l,\alpha_i}$ as: 
\begin{equation}
P_{l,\alpha_i}= P_{l,\alpha_{i-1}} \exp\left(  \tau_i \cdot \bm{m}  \right)
\label{Ps}
\end{equation}
where time is scaled in $N_e = \sum_{d\in D} \theta_d $ generations, the exponential is a matrix exponential, and $\bm{m}$ is the matrix of instantaneous backward migration rates with diagonal elements set such that rows sum up to 0.
If lineages $l_1$ and $l_2$ coalesce to an ancestral lineage $l$ at time $t$, then 
\begin{equation}
P_{l,t} = \dfrac{ \left( \dfrac{ P_{l_1,t,1} P_{l_2,t,1} }{ \theta_1 } , \ldots , \dfrac{P_{l_1,t,|D|} P_{l_2,t,|D|} }{ \theta_{|D|} } \right) }{  \sum_{d=1}^{|D|} \dfrac{ P_{l_1,t,d} P_{l_2,t,d} }{ \theta_d } } 
\label{Ps2}
\end{equation}
which is the normalised entrywise product (element by element product) of the distributions of the coalescing lineages, where $P_{l,t,d}$ is entry $d$ of vector $P_{l,t}$.

Lastly, the contribution to the likelihood of the event at time $\alpha_i$ is:
\begin{equation}
E_i = \left\{  \begin{array}{lcl} 1 & \mbox{if it is a sampling event,} \\ \sum_{d \in D} P_{l,\alpha_i,d} P_{l',\alpha_i,d}   \dfrac{1}{\theta_d} & \mbox{if it is a coalescence event of $l$ and $l'$.}  \end{array}  \right.
\end{equation}
The contribution to the likelihood of sampling events might in principle be modified to account for an informative sampling process~\cite{VOLZFROS14}.
Details of how we efficiently calculate these quantities, in particular Eq.~\eqref{rate}, are given in Supplementary Text S1.

\subsection*{Simulations}

We performed simulations under the structured coalescent~\cite{HUDS90,NOTO90}, simulating a neutral coalescent of samples from different locations (also called demes or populations).
Lineages can only coalesce while in the same location, and migration events happen according to a pre-specified backward-in-time instantaneous migration rate matrix.
So to fit the assumptions of DTA, all populations have the same effective population size ($N_d=1$), and effective population size parameters are not estimated.

In the first simulation setting (called ``continents'') we have two locations, with lineages in the same location coalescing freely.
The migration rate matrix was different for each replicate, each time being sampled from the prior used in DTA: the total forward migration rate was sampled from an exponential distribution with mean $\hat m \in \{0.5, 2.0, 5.0\}$;
The two forward migration rates (that from location 1 to 2, $\hat m_{12}$, and that from two to 1, $\hat m_{21}$) were sampled independently from $\Gamma(1.0,1.0)$ distributions.
The forward migration rates then were scaled such that the total migration rate (as defined in DTA: $\dfrac{2\hat m_{12}\hat m_{21}}{\hat m_{12}+\hat m_{21}}$) was $\hat m$.
Lastly, the instantaneous backward migration rates used to simulate the coalescent process were defined as $m_{12}=\hat m_{21}$ and $m_{21}=\hat m_{12}$.
This favours DTA by matching its prior on forward migration rates, but not the one of MTT or BASTA, where backward migration rates priors are log-normal distributions with $\mu=\ln (\hat m)$ and $\sigma=4$.

In every simulation the structured coalescent process was simulated starting from the leaves (all samples are contemporaneous) and iteratively sampling waiting times for events, each of which is either a coalescence or a migration, according to the distribution $\exp(\sum_{d \in D} \left[ \binom{n_d}{2} +n_d \sum_{d'\neq d} m_{d d'} \right ])$ where $n_d$ is the number of extant lineages in population $d$. 
This process was repeated until the root is reached.

In a second simulation setting (called ``archipelagos'') we have eight locations subdivided in two groups, each containing four locations.
Each location represents an island, while each of the two groups is an archipelago, so that migration rate within an archipelago is high, while it is low between archipelagos.
All migration rates between locations in the same group are identical ($m_1$) as are all migration rates between locations in different groups ($m_2$).
We fixed $m_1/m_2 = 10$, and sampled $m_2$ from an exponential distribution with mean $0.5$.

From all simulations, migration rates and root location were then estimated using DTA~\cite{LEMEETAL09, EDWAETAL11}, MTT and BASTA, all as implemented in BEAST2.
Genetic/phylogenetic information provided in input to the three methods was generated from the simulated coalescent histories in 3 different ways:
in the first setting (``fixed tree''), we assume abundant genomic information, such that the coalescent tree is perfectly known and only the total tree height scale in coalescent units is unknown, which was achieved by fixing the phylogeny in BEAST2 to the simulated one, and only estimating the total tree height;
in a second strategy (``no data''), we assume that genetic data is extremely scarce, and provided no genetic/phylogenetic information at all (this is not realistic, but is useful to learn about prior biases of different models);
lastly, in the third strategy (``variable tree'') we provided an alignment of 2000 bp generated from the simulated coalescent tree using SeqGen~\cite{RAMBGRAS97} with a transition/transversion ratio of $\kappa=3$ and mutation rate of 0.01 in units of $N_e$ generations, while we estimated the phylogenetic tree in BEAST2 along with the phylogeographic parameters.

\subsection*{Ebola Transmission Study}

To study changes of host type in Ebola we used whole genome Ebola sequences from 78 patients recently obtained and aligned with sequences from previous outbreaks~\cite{GIREETAL14}.
The authors of this study investigated the phylogenetic relationship of samples within or between Ebola outbreaks.
Here, we apply the three phylogeographic methods presented above to infer the contribution of zoonotic events to Ebola spread. 
We used the same alignment provided in~\cite{GIREETAL14} for the BEAST analysis, including sampling dates, but we also added information regarding host type.
All samples were specified to be from human hosts, but despite this, we allowed lineages to switch to an animal reservoir during their history.
Migration was only allowed from animal to human host, and not vice-versa.
So in our phylogeographic model we had two locations (respectively human and animal reservoir) but migration was only assumed to occur in one direction.
This results in a structured coalescent model with three parameters for MTT and BASTA (one migration rate and two effective population sizes), but only two parameters for DTA, as only a single general effective population size can be defined in that model.
A peculiarity of this analysis is that only genetic samples from one of the two considered populations were available.
While this might seem an impassable limitation, previous studies have shown that the structured coalescent provides meaningful estimates even in the absence of samples from one populations (i.e. ``ghost deme'', see~\cite{EWINRODR06}), suggesting that it is possible to perform statistical inference of zoonosis rates from this dataset.

\section*{Results}

\subsection*{DTA is Inherently Biased by the Sampling Process}

The introduction of approximation can lead to biases in parameter estimation.
To test for the presence of biases associated with sampling strategy in DTA, MTT and BASTA, we performed simulations paired with estimation in the scenario where the genetic data was completely uninformative.
In the absence of informative genetic data, such as in our ``no data'' scenario, the posterior distribution of a well calibrated Bayesian approach should coincide with the prior distribution, and we want to test if this is the case here.
One of the differences between DTA and the structured coalescent is that the sampling locations in the structured coalescent are auxiliary variables which are always conditioned over, while in DTA they are treated as part of the data.
Usually, sampling locations should not be treated in the same way as genetic data because they may be quite arbitrary.

\begin{figure}
\caption{Figure S1}
\includegraphics[width=0.99\textwidth]{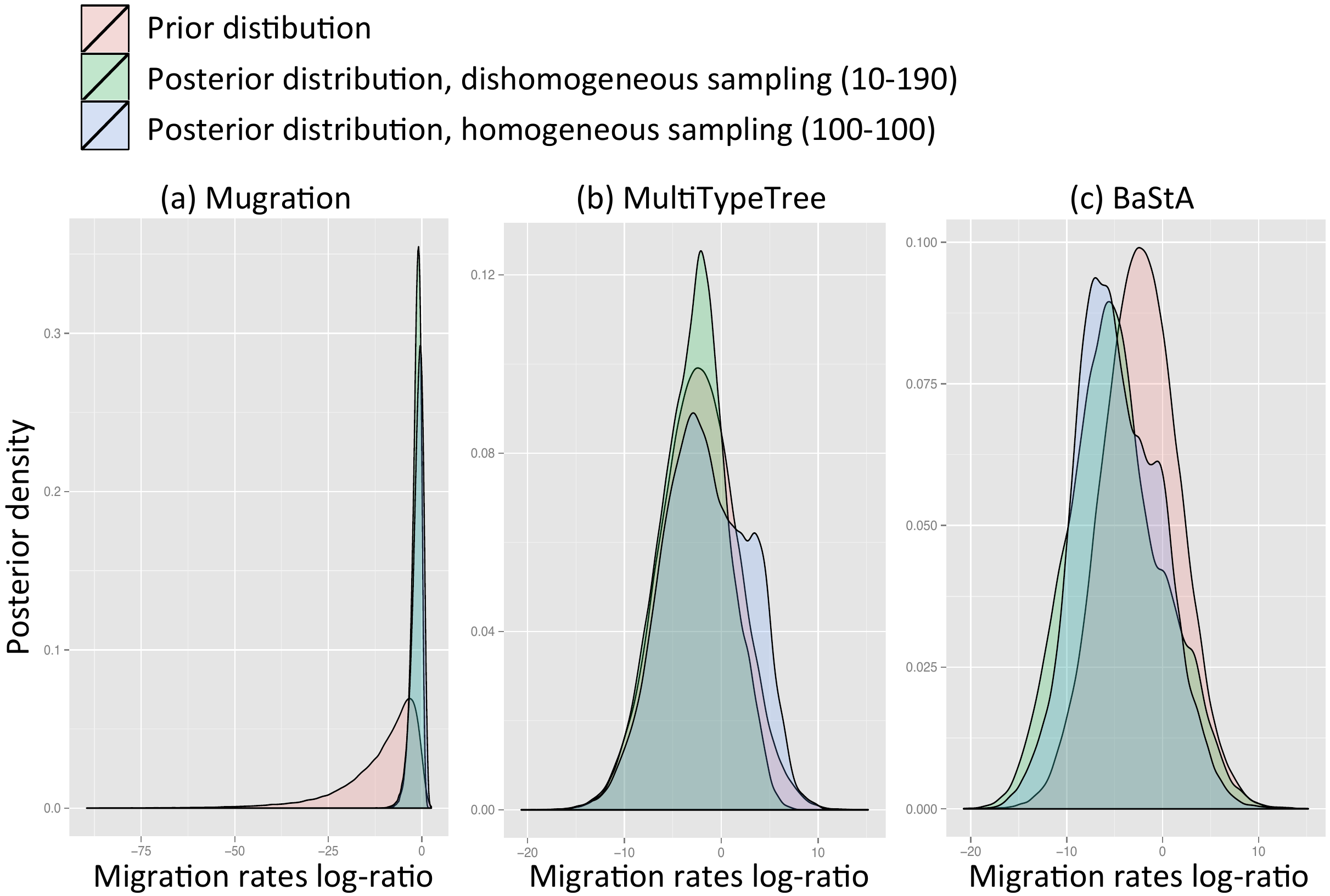}
\label{FS1}
\end{figure}

Simulations show that the posterior inference of DTA depends heavily on the distribution of sampling locations, unlike the two structured coalescent methods.
Particularly with high migration rates (prior mean 5.0) DTA posteriors show large biases (Fig.~\ref{noData}A), meaning that the sampling strategy chosen significantly influences the inference. 
The posterior distributions of the other two methods are noticeably less smooth (Fig.~\ref{noData}B and C), reflecting the larger computational demand of MTT and BASTA.
When migration rates are low (prior mean 0.1) DTA over-estimates them (\nameref{FS1}A).
This is because sampling from two locations implies the presence of at least one migration event.
However, under the DTA model there is a high prior probability of no migration when migration rates are sufficiently low. 
In contrast, the structured coalescent accounts for the fact that there must be at least $D-1$ migration events when $D$ locations are sampled.

\begin{figure}[h]
\caption{{\bf DTA is inherently biased by the sampling process.} 
To show this, we considered the case of a dataset with totally uninformative sequences (this mimics extremely short or uniform sequences).
DTA treats the sampling process as informative about migration parameters, unlike the structured coalescent.
This can be seen by the perturbation of the posterior relative to the prior distribution, as inferred by (a) DTA, (b) MTT, (c) BASTA.
We used high migration rates, with prior mean 5.0, 
and two sampling strategies: homogeneous (100 samples per locations, azure) and inhomogeneous (respectively 10 and 190 samples per location, green).
In pink is the prior distribution.
On X axis is the ratio of the two migration rates, on Y axis the density of the corresponding value in the distribution.
Each plot is obtained from 10 merged posteriors of independent MCMC runs each of $5\times 10^6$ iterations.}
\label{noData}
\end{figure}

\subsection*{DTA Under-represents Uncertainty}

Next we addressed the accuracy of the methods when dealing with highly informative sequences.
At the opposite extreme to before, methods are expected to perform best when genetic information is so informative that the phylogenetic tree can be estimated with great detail.
When genetic data are extremely informative, for example in some cases when whole genomes are available, it may be convenient to assume that the phylogeny is known.
Here we consider this scenario by providing the true tree topology and relative branch lengths as input, and estimating only the tree height together with the migration rate parameters. 

\begin{figure}
\caption{Figure S2}
\includegraphics[width=0.99\textwidth]{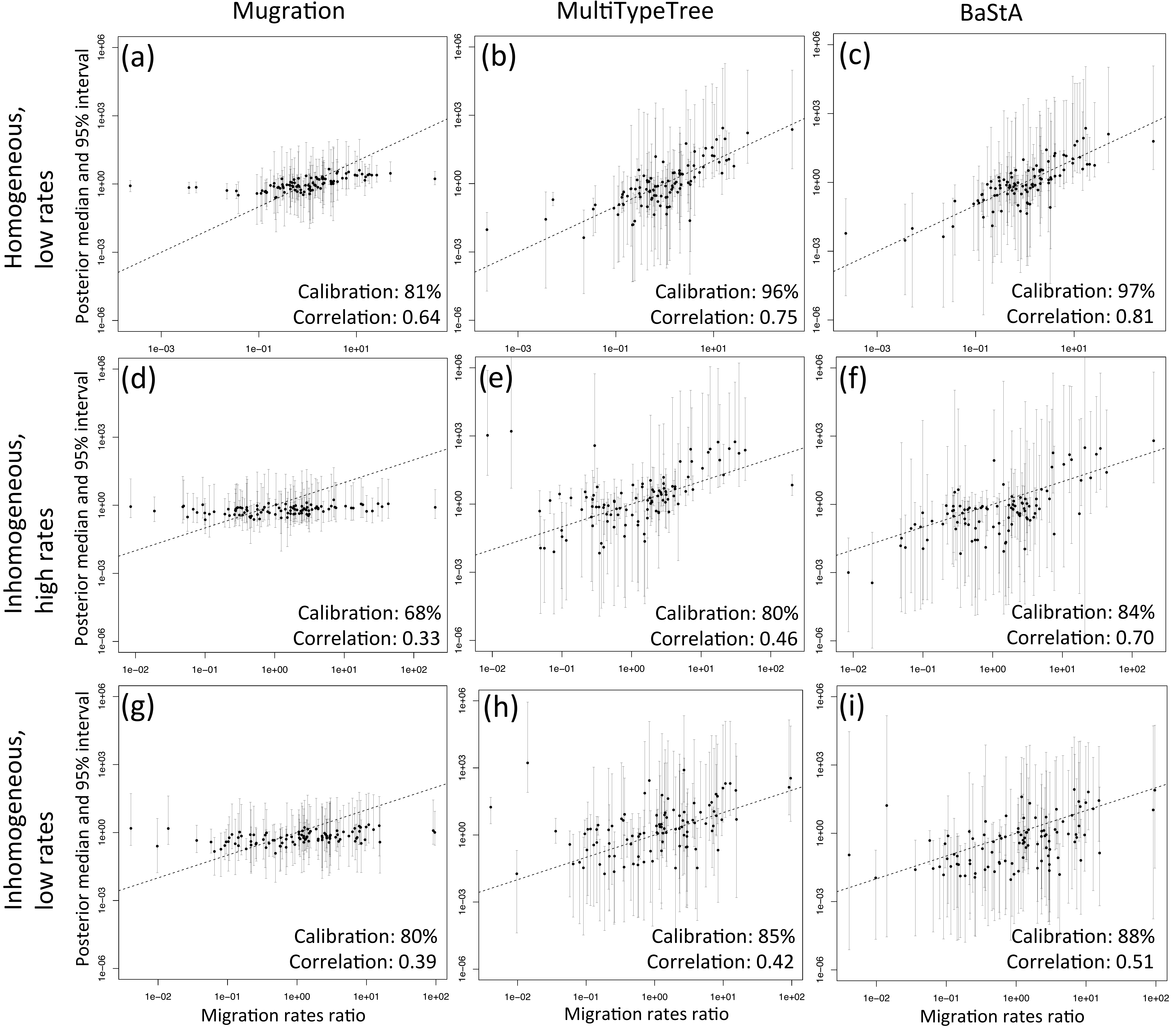}
\label{FS2}
\end{figure}

\begin{figure}
\caption{Figure S3}
\includegraphics[width=0.99\textwidth]{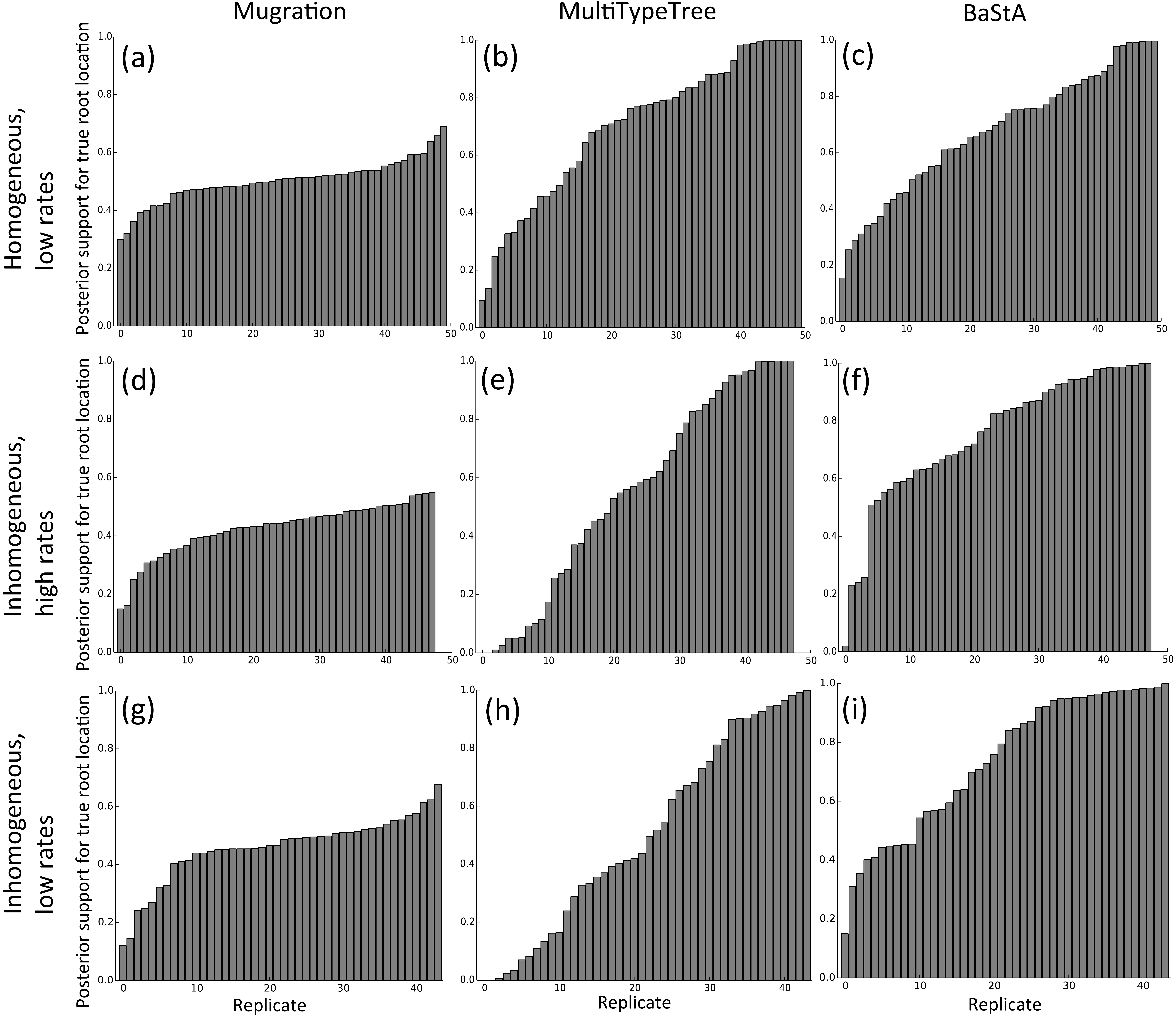}
\label{FS3}
\end{figure}

The absolute migration rates have different scales under DTA and the structured coalescent, so we focus on estimation of the relative migration rate, the ratio of the two considered migration rates.
DTA exhibits generally poor performance (Fig.~\ref{allDataRates} and \nameref{FS2}).
The $95\%$ confidence intervals are not well calibrated (they include the truth $56\%$-$81\%$ of the time, compared to $80\%$-$96\%$ for MTT and $84\%$-$97\%$ for BASTA; Table~\ref{allDataTable}), and the correlation between simulated and estimated values is lower (0.33-0.64) than competing methods (0.51-0.85 for BASTA and 0.42-0.77 for MTT; Table~\ref{allDataTable}).
Estimation of root location shows similar trends (Fig.~\ref{allDataRoot} and \nameref{FS3}).
Earlier structured coalescent methods had increased computational demands with elevated migration rates due to the increased parameter space needed to represent migration histories~\cite{EWINETAL04}.
Here we found that MTT performed similarly well with different total migration rates, supporting the view that its new proposal functions represent a very considerable improvement over previous efforts~\cite{VAUGETAL14}.

\begin{figure}
\caption{Figure S4}
\includegraphics[width=0.99\textwidth]{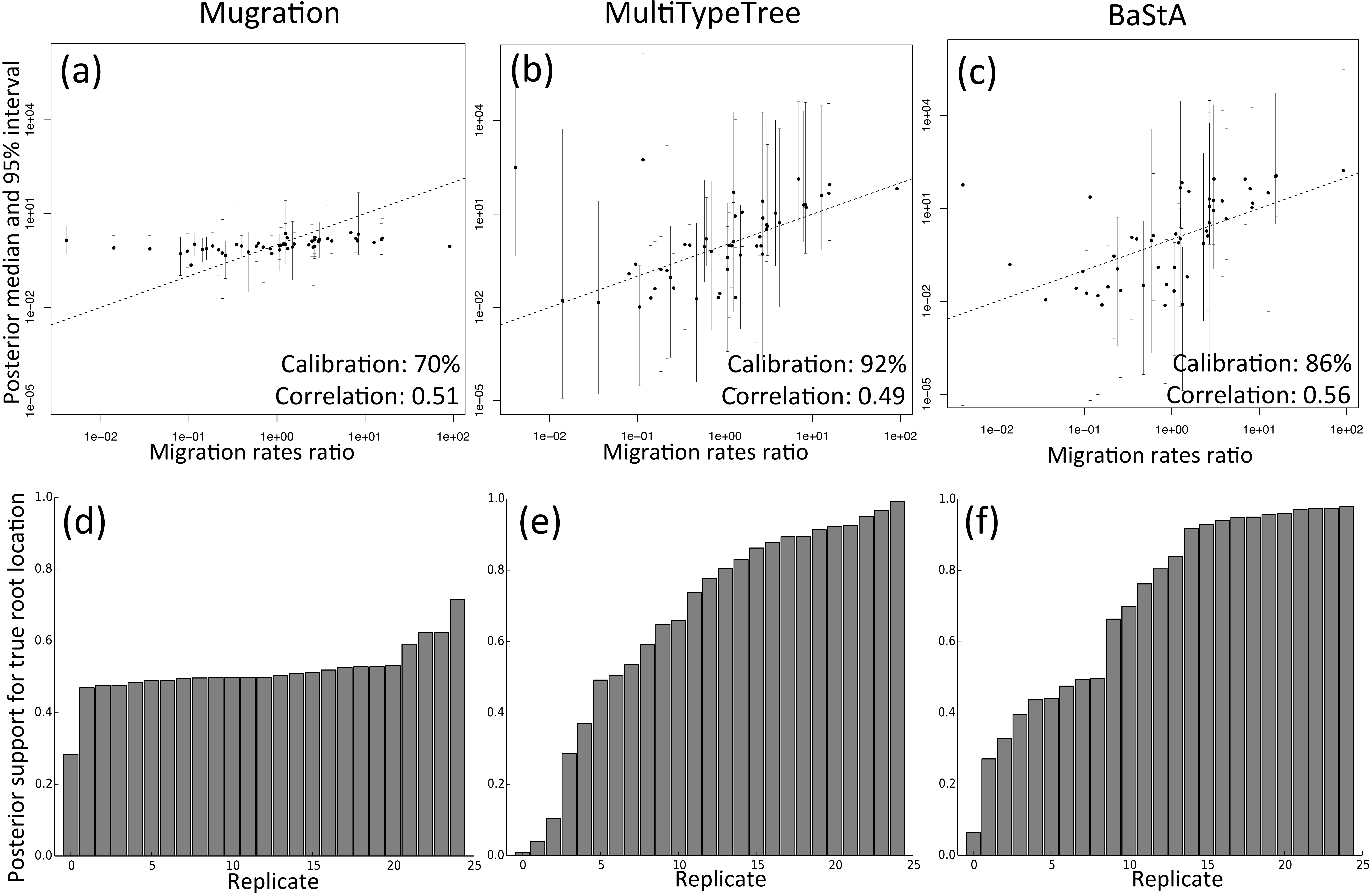}
\label{FS4}
\end{figure}

We also compared the performance of the different approaches at intermediate levels of information, when there is both phylogenetic signal and phylogenetic uncertainty (the ``variable tree'' scenario).
This scenario is probably the most common in practice, but also the most complex as phylogenetic uncertainty makes inference more computationally demanding.
All three methods account for phylogenetic uncertainty by exploring the space of possible trees with MCMC.
To investigate scenarios with phylogenetic uncertainty, we simulated alignments of 2000 bp with a mutation rate of 0.01 per $N_e$ generation.
We simulated 50 replicates and a single scenario (prior total migration rate of 2.0, and 50 samples per population).
All methods reported greater uncertainty in this setting, as expected, with DTA showing weaker correlation between point estimates and the truth and severely underestimating posterior uncertainty compared to BASTA. 
While MTT most faithfully captures posterior uncertainty, it shows the worst correlation between point estimates and the truth, possibly reflecting its greater computational demands in the presence of phylogenetic uncertainty (\nameref{FS4} and Table~\ref{varTreeTable}).

\begin{figure}[h]
\caption{ {\bf DTA retrieves partial information and is not calibrated.}
We simulated the scenario of complete phylogenetic information, that is, when the phylogenetic tree is perfectly known to each of the methods.
In this scenario methods are expected to give the best accuracy.
We plot the estimates of the ratio of the two migration rates between the two simulated populations using (a) DTA, (b) MTT, (c) BASTA.
Furthermore, for each method we show the Pearson correlation between the true value and the posterior median (``Correlation''), and the proportion of replicates in which the true value lies in the $95\%$ confidence interval (``Calibration'').
We simulated high migration rates (prior mean 5.0), homogeneous sampling (100 samples per locations), and 100 replicates.
The simulated (true) ratio of the two migration rates is on the X-axis, while the estimated ratio is on the Y-axis.
The diagonal dashed line represents the hypothetical perfect estimate.
Each dot represents a posterior median, and intervals show the posterior $95\%$ coverage.
Number of MCMC steps for DTA, MTT and BASTA are respectively $10^6$, $2\times 10^5$ and $10^5$ so to achieve similar running times (respectively approximately 180, 200 and 150 seconds per replicate).}
\label{allDataRates}
\end{figure}

\begin{table}[!ht]
\caption{\textbf{Summaries of simulations with two populations in the ``fixed tree'' scenario. }}
\begin{tabular}{l l l *{3}{c}}
Sampling & Rate & Method & Calibration & Correlation & RMSE \\
\hline
Homogeneous & Fast & DTA & 0.56 & 0.58 & 1.83  \\
 & & MTT & 0.87 & 0.77 & 1.32  \\
 & & BASTA & 0.95 & 0.83 & 1.51  \\
\hline
Homogeneous & Slow & DTA & 0.81 & 0.64 & 1.65   \\
    &  & MTT & 0.96 & 0.75 & 1.52   \\
    &  & BASTA & 0.97 & 0.81 & 1.30   \\
\hline    
 Inhomogeneous   & Fast & DTA & 0.68 & 0.33 & 1.79   \\
    &  & MTT & 0.80 & 0.46 & 2.50   \\
    &  & BASTA & 0.84 & 0.70 & 2.08   \\
\hline    
 Inhomogeneous & Slow & DTA & 0.80 & 0.39 & 1.73   \\
    &  & MTT & 0.85 & 0.42 & 2.49  \\
    &  & BASTA & 0.88 & 0.51 & 2.29   \\ \hline
\end{tabular}
\begin{flushleft} For each scenario 100 replicates were performed. 
For each summary the value of interest is the logarithm of the ratio of the two migration rates.
``Sampling'' refers to the sampling strategy adopted: ``Homogeneous'' means that 100 samples per populations were used, ``Inhomogeneous'' means that we used 10 samples for one population and 190 for the other.
``Rate'' refers to the total migration rate: ``Fast'' represents a prior mean of 5.0, while ``Slow'' a prior mean of 0.5.
``Calibration'' represents the proportion of replicates for which the true value falls within the $95\%$ posterior interval.
``Correlation'' represent the Pearson correlation between the simulated value and the posterior median.
``RMSE'' represent the root mean square error of the posterior median.
\end{flushleft}
\label{allDataTable}
\end{table}

\begin{figure}[h]
\caption{{\bf Structure coalescent improves ancestral location inference.}
We measured the accuracy in inferring the location of phylogeny roots for different methods: (a) DTA, (b) MTT, (c) BASTA.
Each bar represents the posterior support for the true root location for a replicate, so taller bars represent better inference.
Simulations were performed with two locations, fixed trees, high migration rates (prior mean 5.0), and homogeneous sampling (100 samples per locations).
For every scenario there are 100 replicates, but only replicates for which the simulated root is in the first location are shown.
Bars are in increasing order.
Number of MCMC steps for DTA, MTT and BASTA are respectively $10^6$, $2\times 10^5$ and $10^5$ so to achieve similar running times (respectively approximately 180, 200 and 150 seconds per replicate).}
\label{allDataRoot}
\end{figure}

\begin{table}[!ht]
\caption{{\bf Summaries of simulations with two populations and variable trees. }}
\begin{tabular}{l *{3}{c}}
Method  & Calibration & Correlation & RMSE \\
\hline
DTA & 0.70 & 0.51 & 1.68  \\
MTT & 0.92 & 0.49 & 2.56  \\
BASTA & 0.86 & 0.56 & 2.61  \\
\hline
\end{tabular}
\begin{flushleft} The datasets are simulated under moderate migration rates (prior mean 2.0), homogenous sampling (50 samples per location), with a total of 50 replicates.
For each summary the value of interest is the logarithm of the ratio of the two migration rates. 
``Calibration'' represents the proportion of replicates for which the true value falls within the $95\%$ posterior interval.
``Correlation'' represent the Pearson correlation between the simulated value and the posterior median.
``RMSE'' represent the root mean square error of the posterior median.
\end{flushleft}
\label{varTreeTable}
\end{table}

\subsection*{BASTA is Faster than Structured Coalescent Methods}

\begin{figure}
\caption{Figure S6}
\includegraphics[width=0.99\textwidth]{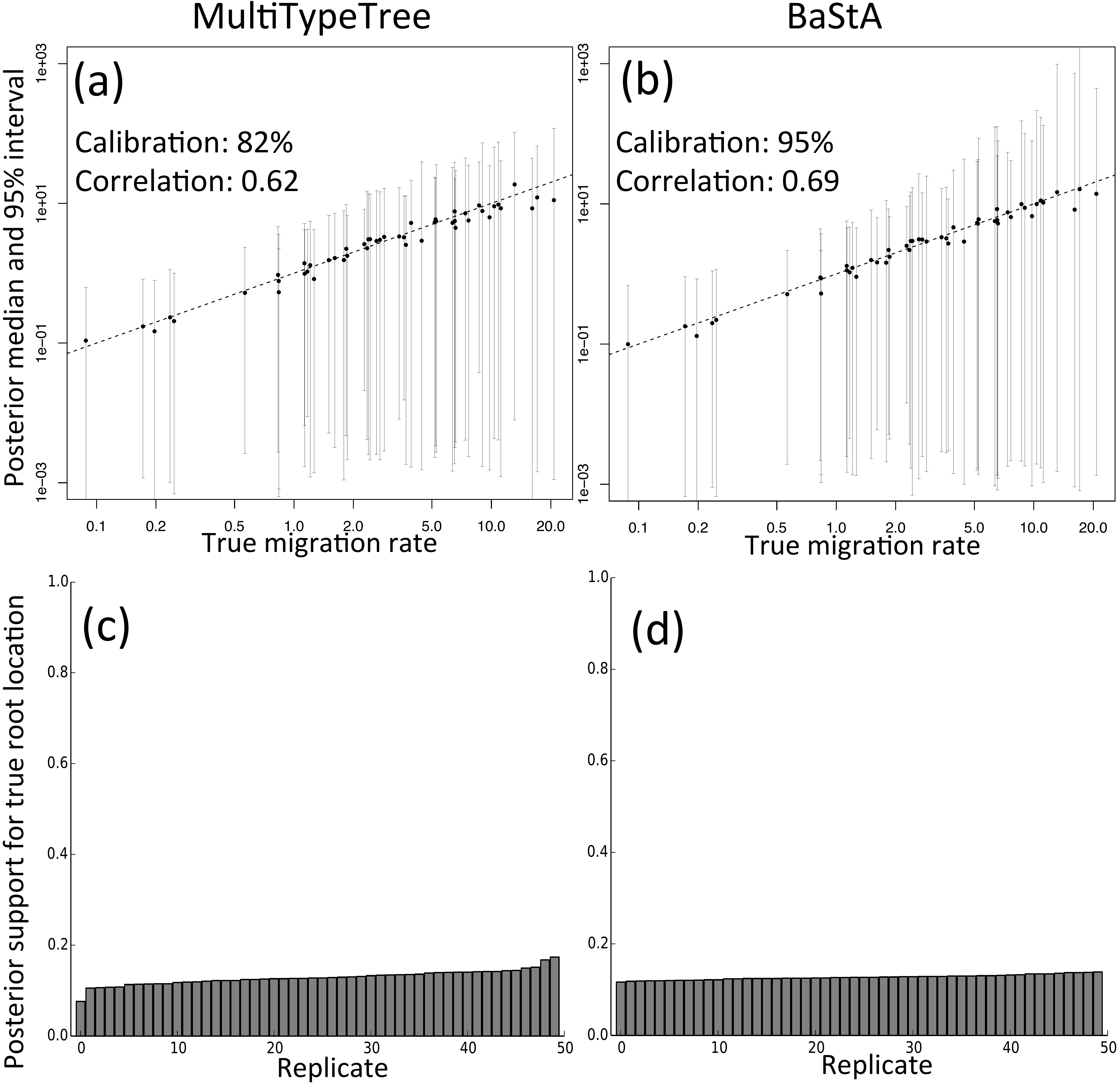}
\label{FS6}
\end{figure}

So far, we have considered scenarios with only two populations, for which structured coalescent models are expected to work well.
Yet, with more populations, structured coalescent methods can be too computationally demanding for practical inference.
To test the performance of BASTA in such situations, and compare it to MTT, we simulated a scenario with eight populations and fixed the tree to facilitate inference under the structured coalescent.
We subdivided the populations into two archipelagoes (clusters of islands), each of four islands, and with 40 samples from each island.
Migration between islands in the same archipelago was fast (mean 5.0) while migration between archipelagoes was 10-fold lower.
Both methods reported considerable uncertainty in their estimates of the migration rates and root location (\nameref{FS6}).
However, BASTA reached convergence and acceptable effective sample sizes in reasonable time ($2\times 10^6$ MCMC steps for a total of approximately $1.3\times 10^4$ seconds per chain);
instead, with similar computational effort, MTT was far from convergence (see e.g. Fig.~\ref{trace} and~\nameref{FS6} for some randomly sampled replicates) with most parameters having an effective sample size (ESS, measuring how well the posterior has been explored, see~\cite{KASSETAL98}) below 20, way below the generally accepted limit of 200.
These results show that not only does BASTA produce a modest but consistent improvement in calibration and informativeness over MTT (see also Table~\ref{8popTable}) but has also broader applicability to scenarios with more populations.

\begin{figure}[h]
\caption{{\bf BASTA has broader applicability than MTT.}
When simulating a scenario with eight populations, BASTA always seemed to efficiently explore the parameter space in acceptable time, while MTT, with comparable computational resources, never achieved convergence.
With these plots we show the traces of the posterior probability (Y axis) over the MCMC steps (X axis) in one random replicate.
Similar plots for further replicates are found in~\nameref{FS6}. 
For these simulations we used fixed trees.}
\label{trace}
\end{figure}

\begin{table}[!ht]
\caption[]{\textbf{BASTA improves rate estimation.}}
\begin{tabular}{l l *{5}{c}}
Method  & Calibration & Correlation & RMSE \\
\hline
within archipelago&&\\
\hline
MTT & 0.815 & 0.62 & 1.49  \\
BASTA & 0.95 & 0.69 & 1.33  \\ \hline
between archipelagos&&\\
\hline
MTT & 0.98 & 0.61 & 1.57  \\
BASTA & 1.00 & 0.67 & 1.47  \\ \hline
\end{tabular}
\begin{flushleft} To compare migration rate estimation between MTT and BASTA in a setting with many populations, we simulated a scenario with two groups of populations (two archipelagos) each containing four populations (four islands) and with 40 samples per population. 
50 replicates were simulated.
The values of interest considered here are the logarithm of the migration rate between islands in the same archipelago (top part of the table) and between islands in different archipelagos (bottom part of the table). 
``Calibration'' represents the proportion of rates among all replicates for which the true value falls within the $95\%$ posterior interval.
``Correlation'' represent the Pearson correlation between the simulated value and the posterior median.
``RMSE'' represent the root mean square error of the posterior median.
\end{flushleft}
\label{8popTable}
\end{table}

\subsection*{Model Choice Strongly Influences Reconstruction of Ebola Transmission Dynamics}

While we write, the most deadly known outbreak of Ebola virus is ongoing in West Africa.
In recent work, Gire et al.~\cite{GIREETAL14} have collected and whole genome sequenced 99 Ebola virus samples from 78 patients.
Using these and previous data, the authors have shown that all available sequences within each outbreak since 1976 cluster together phylogenetically;
furthermore, divergence of lineages leading to different outbreaks usually considerably pre-dates the older outbreak.
This fact and the shape of the inferred phylogeny suggest that of independent zoonotic transmissions are the source of different Ebola outbreaks in humans.
Ebola infections in different animals have been directly observed more than 50 times, with bats thought to be the main reservoir~\cite{PIGOETAL14}.

We addressed this subject in order to explore the potential impact of modelling considerations on epidemiological conclusions based on genetic data.
We defined a highly simplified phylogeographic model with two subpopulations: the first representing human hosts, the second representing an animal reservoir.
In this model, coalescence events within the human population represent human-to-human transmission; 
similarly coalescence events in the animal reservoir represent transmission between animal hosts. 
Migration from the animal reservoir to the human population corresponds to a zoonotic transmission.
Migration from human to animal was assumed sufficiently rare to be ignored (see~\cite{PIGOETAL14}).

Using this phylogeographic model, we investigated the effect of model choice - DTA versus structured coalescent - on the epidemiological conclusions concerning the role of zoonotic transmission in seeding human outbreaks of Ebola. 
We found that the two models gave diametrically opposed results.

Consistent with general understanding of the emergence of Ebola outbreaks in humans, the structured coalescent models, implemented in MTT, inferred that each outbreak was seeded by an independent zoonosis from the Ebola reservoir population (Fig.~\ref{Ebola}a), with 18 animal-human zoonoses inferred in the history of the sample ($95\%$ CI $[15.0,22.0]$), from 1976 until present. 
In keeping with this, the effective size of the animal reservoir was inferred to be very large ($95\%$ CI $[7.3,22.9]$) compared to the effective size of the virus population sustained in humans ($95\%$ CI $[0.27,0.60]$ ). 
The most recent common ancestor of all sampled human outbreaks was inferred to have originated in the animal reservoir population with $100\%$ posterior probability.
These results were also supported by BASTA.

In direct contrast, the DTA painted a very different picture of Ebola outbreak emergence that does not accord with scientific understanding. 
With high confidence, no zoonotic transmissions from animals to humans were inferred in the history of the sampled outbreaks ($100\%$ posterior probability), with the most recent common ancestor inferred to have occurred in the human population. 
Despite the implausibility of undetected human outbreaks having sustained Ebola virus in humans over four decades, the discrete trait model supported this scenario with high confidence.

These results illustrate the strong influence of model choice on phylogeographic inference. 
They demonstrate the possibility of obtaining implausible results supported by high confidence with DTA. 
Although in the case of Ebola, the strength of evidence concerning the epidemiology of the disease is more than sufficient to disregard the discrete trait analysis out of hand, it demonstrates the potential to produce highly misleading inference when independent epidemiological understanding is scarce.

\begin{figure}[h]
\caption{{\bf Different histories are inferred by different methods.}
The trees were inferred using (a) DTA and (b) MTT. In red we show regions of the phylogeny inferred to be in human host, while in blue are regions inferred to be in animal reservoirs.
The scale of the axis is in number of years from present. Sampling dates and locations are included in sample names.
Part (a) shows the maximum clade credibility tree with median node height inferred with DTA, while part (b) displays a tree sampled from MTT output.}
\label{Ebola}
\end{figure}

\section*{Discussion}

Phylogeography has rapidly gained prominence in a wide range of settings where it can quantify historical patterns of migration from genetic data alone.
In the context of infectious disease epidemics, phylogeographic methods can infer transmission rates and patterns of spread even in the complete absence of reliable epidemiological information (see e.g.~\cite{LEMEETAL09, LEMEETAL10, MAYETAL11, ALLIETAL12}).
Yet, these methods have only partially been tested and compared.
Here, with simulations based on explicit process-driven population genetics-based models, we showed that different methods exhibit dramatic differences in their inference properties.

While discrete trait analysis (DTA) is extremely fast and accounts for phylogenetic uncertainty, it has problems estimating the correct migration rates even with as few as two populations.
In particular, DTA is sensitive to the relative sampling intensity of populations, such that the sampling strategy adopted can influence the results, particularly when migration rates are high and genetic data are sparse.
We reiterate that here we have assessed the performance of DTA as a model of migration, and not in the context of evolution of discrete traits, such as some phenotypes, for which the discrete trait analysis is appropriate.
MTT, on the other hand, proved better calibrated, with less biased estimates, with a stronger correlation between simulated and estimated values.

Together with other methods based on structured coalescent, MTT also has the advantage over DTA of being able to estimate and account for differences in population sizes between populations.
Also, it provides estimates of absolute migration rates that are meaningful from a population genetics perspective.
We want to acknowledge that MTT proved useful even in contexts of elevated migration rates, where previous structured coalescent-based methods showed convergence problems.
Yet, we also show that when several demes are considered (we simulated eight demes, but~\cite{VAUGETAL14} suggest not to exceed four) MTT can have convergence issues.
To deal with this problem, we propose a new approach, BASTA, based on an approximation of the structured coalescent similar to some recently proposed~\cite{VOLZ12, RASMETAL14}.
BASTA integrates over all the possible migration histories rather than sampling them, therefore considerably reducing the parameter space that needs to be explored.
Not only did this approach show appreciable improvements in accuracy with respect to MTT with just two populations, but it was possible to obtain MCMC convergence with BASTA for as many as eight populations, which was beyond the reach of MTT in feasible time (3-4 hours).

Finally, using real data from Ebola outbreaks in humans we showed that the choice of model is very important, because different models can lead in practice to completely different results.
In fact, diametrically opposite phylogeographic patterns were estimated using DTA rather than structured coalescent-based methods.
MTT and BASTA gave instead comparable estimates.
We recommend that users exercise caution, and we point out that methods based on the structured coalescent are in general more reliable, although also more computationally demanding.
The fact that all three approaches considered here are all implemented in the same phylogenetic package (BEAST2) is a considerable advantage, as it is possible to run and compare different methods while installing a single piece of software and using similar formats.

In the future, we will work on extending BASTA by including estimates of ancestral locations of internal nodes other than the root, and estimates of expected numbers of migrations between populations.

\section*{Acknowledgments}
We thank Erik Volz for the insightful discussion and suggestions on the BASTA approximations.
We also thank Timothy Vaughan for the help with MultiTypeTree, on which BASTA is strongly based.

\nolinenumbers

%
%
%





\newpage

\section*{Supplementary Text S1: Computational Details of BASTA - Eq.~\eqref{rate}}

Calculating Eqs. \ref{Ps} and \ref{Ps2} requires similar steps to Felsenstein's pruning algorithm, and also has similar computational demands.
We therefore do not focus on its details here.
Instead we show how we calculate the coalescent rates (Eq.~\eqref{rate}), and in particular, the sum 
\begin{equation}
\sum_{d \in D} \sum_{l \in \Lambda}\sum_{l' \in \Lambda, l'\neq l} P_{l,t}^d P_{l',t}^{d} \dfrac{1}{\theta_d}
\label{calc}
\end{equation}
for a given time $t$, a given set of extant lineages $\Lambda$, and given the probabilities $P_{l,t}^{d}$.
For brevity, from now on we ignore the time index $t$.
If the expected number of lineages in a deme $d$ is represented as $\mathbb{E}(n_d) :=\sum_{l \in \Lambda} P_l^d$, we have:

\begin{multline}
\sum_{d \in D} \sum_{l \in \Lambda}\sum_{l' \in \Lambda, l'\neq l} P_{l}^d P_{l'}^{d} \dfrac{1}{\theta_d}=  \\
\sum_{d \in D} \left[ \left( \sum_{l \in \Lambda}\sum_{l' \in \Lambda} P_{l}^d P_{l'}^{d} \dfrac{1}{\theta_d} \right) - \left( \sum_{l \in \Lambda} P_{l}^d P_{l}^{d} \dfrac{1}{\theta_d} \right) \right] = \\
\sum_{d \in D}  \dfrac{1}{\theta_d}  \left[ \mathbb{E}(n_d) \mathbb{E}(n_d) -  \sum_{l \in \Lambda} P_{l}^d P_{l}^{d} \right] .
\label{calc2}
\end{multline}

Let us call $S_d= \sum_{l \in \Lambda} P_{l}^d P_{l}^{d}$.
Calculating $S_d$ requires $O(|\Lambda|)$ time and is needed for each deme and twice for each coalescent event.
So, if $n$ denotes the number of samples, the total cost of computing $S_d$ for the whole tree is approximately $O(n^2 \cdot |D|)$.
Updating $S_d$ after a coalescent or sampling event is trivial and negligible in time.
Calculating $\mathbb{E}(n_d)$ is also faster, as we can use the same procedure in Eq.~\eqref{Ps} which avoids the sum over lineages, giving a required time of $\approx O(n \cdot |D|^2)$.
Updating $\mathbb{E}(n_d)$ after coalescence and sampling events is trivial and fast.
Calculating the exponential of the migration rate matrix used in Eq.~\eqref{Ps} is required once per event, for a total computational cost $<O(n \cdot |D|^3)$.
Lastly, while Eq.~\eqref{Ps2} requires negligible time, Eq.~\eqref{Ps} has computational cost $\approx O(|D|^2)$, that repeated over all lineages and over all events, brings to a total cost of $\approx O(n^2 \cdot |D|^2)$, which is the computational bottleneck of BASTA (generally $n>>|D|$).

To further reduce the computational time, we adopt a caching technique that consists in using the same vectors for lineages that have undergone the same history since sampling (including same sampling location).
If many leaves are sampled at the same time, this leads to important savings, but in the worst scenario the total computational demand remains $\approx O(n^2 \cdot |D|^2)$.

\newpage

\begin{figure}
\caption{Figure 1}
\includegraphics[width=0.99\textwidth]{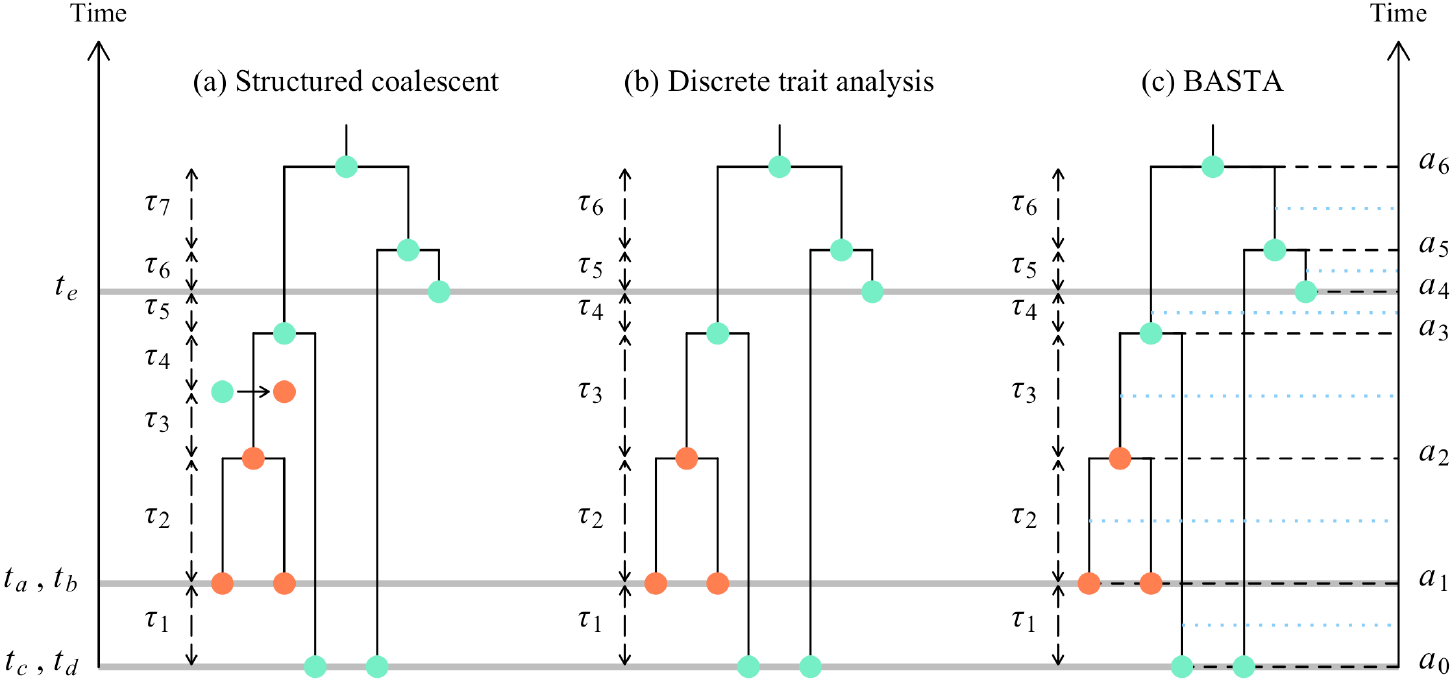}
\label{F1}
\end{figure}

\begin{figure}
\caption{Figure 2}
\includegraphics[width=0.99\textwidth]{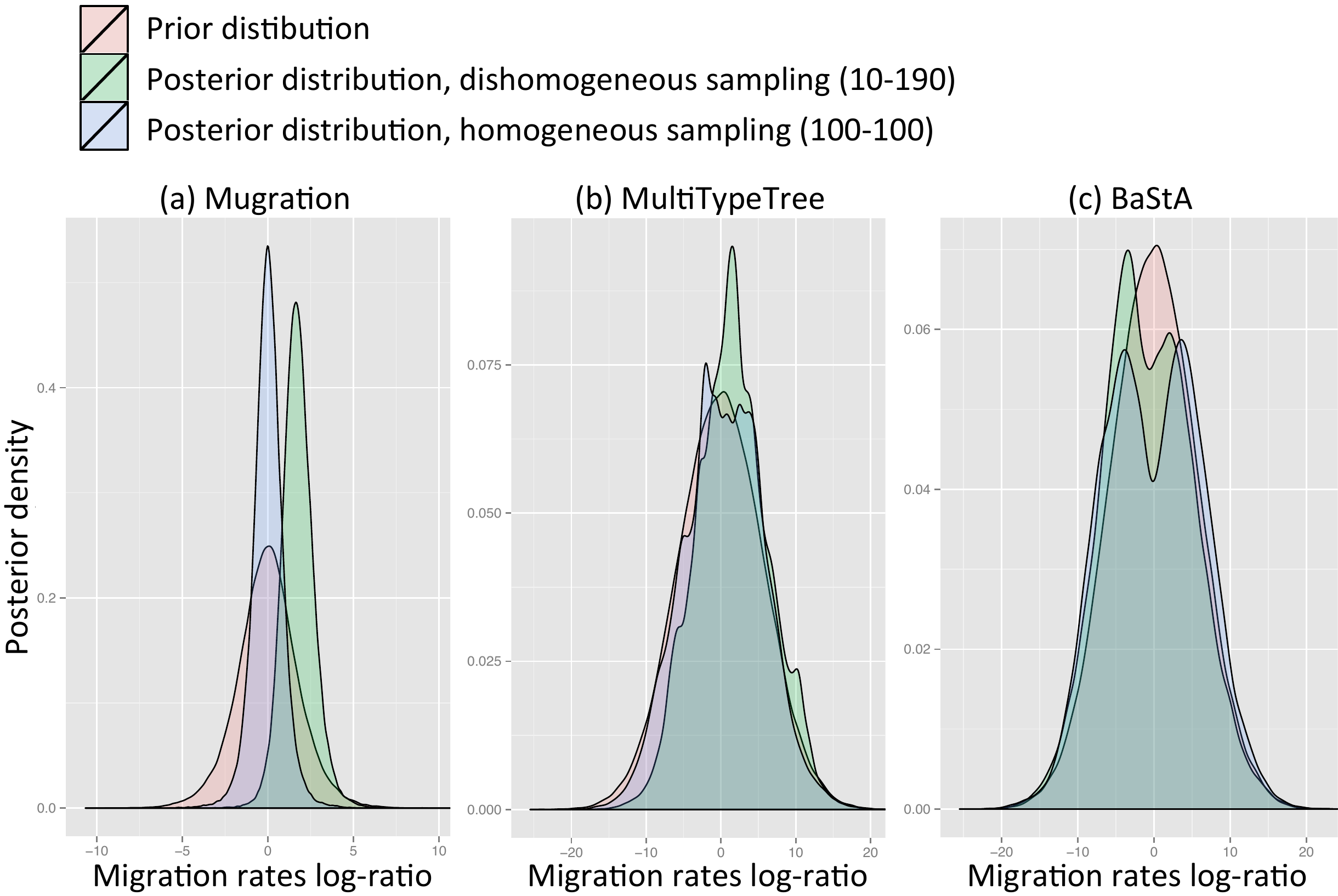}
\label{F2}
\end{figure}

\begin{figure}
\caption{Figure 3}
\includegraphics[width=0.99\textwidth]{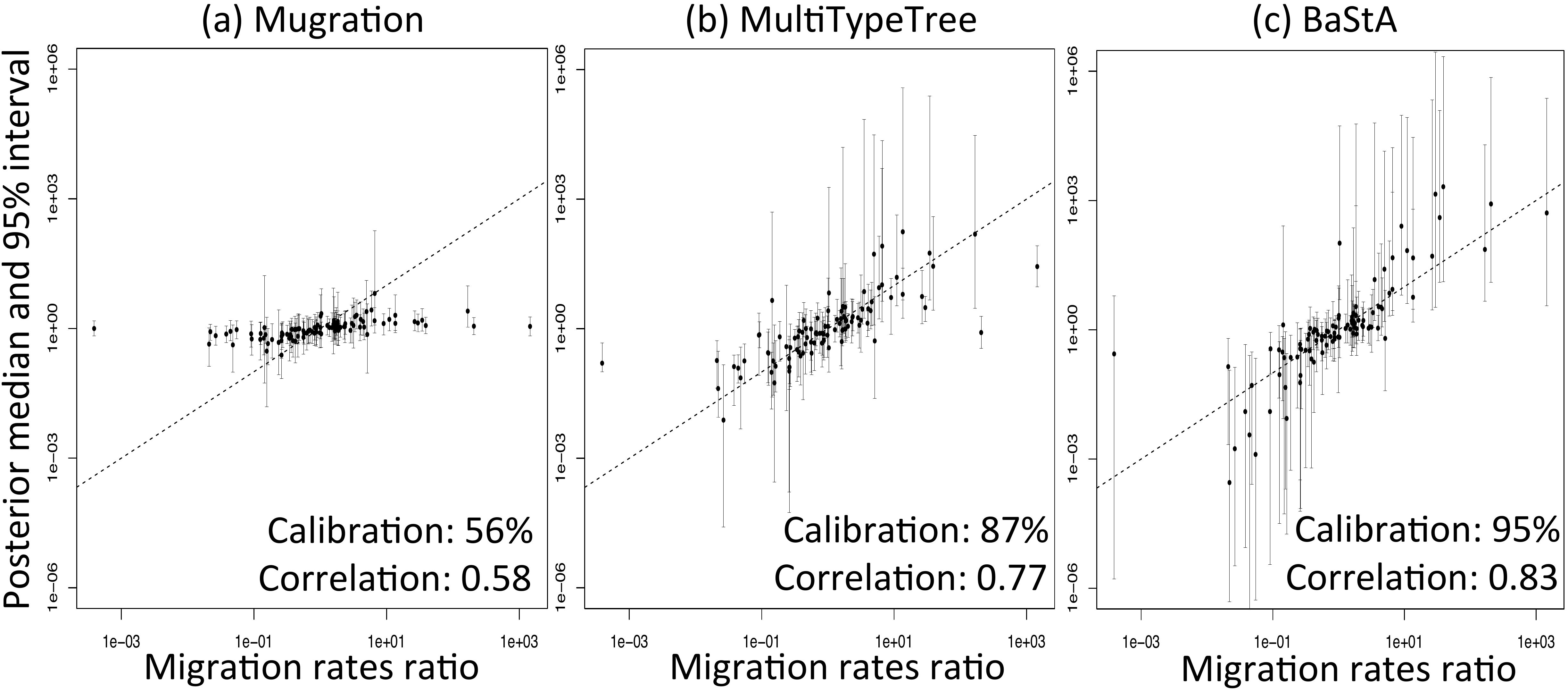}
\label{F3}
\end{figure}

\begin{figure}
\caption{Figure 4}
\includegraphics[width=0.99\textwidth]{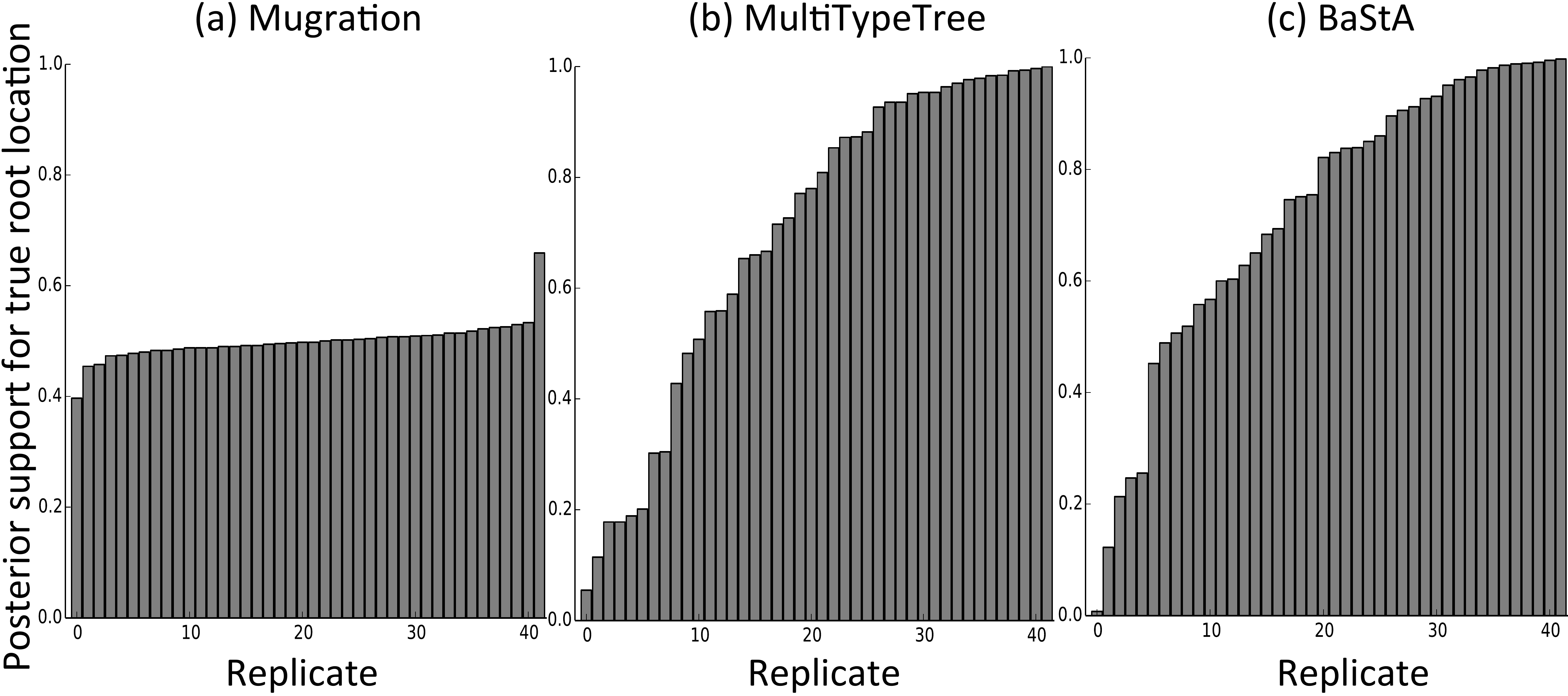}
\label{F4}
\end{figure}

\begin{figure}
\caption{Figure 5}
\includegraphics[width=0.99\textwidth]{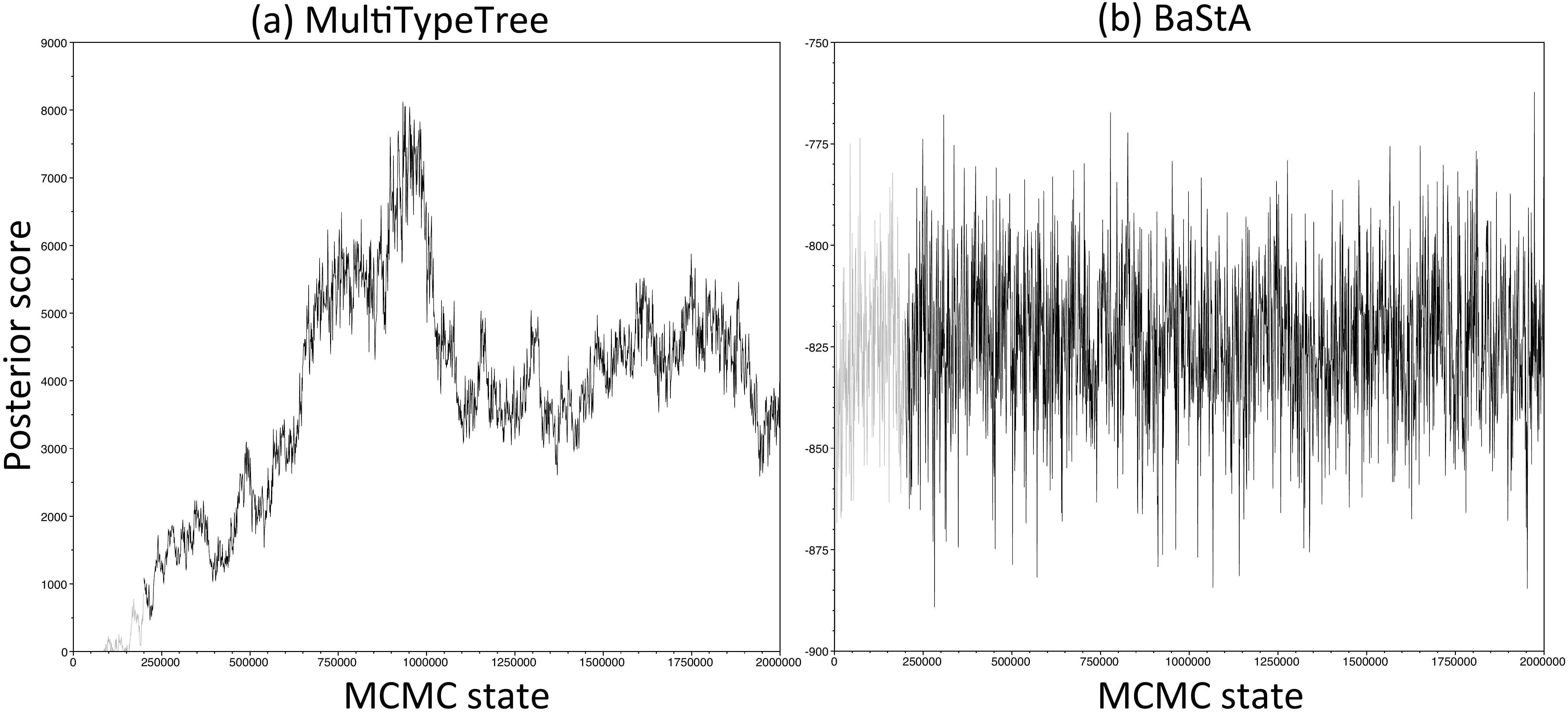}
\label{F5}
\end{figure}

\begin{figure}
\caption{Figure 6}
\includegraphics[width=0.99\textwidth]{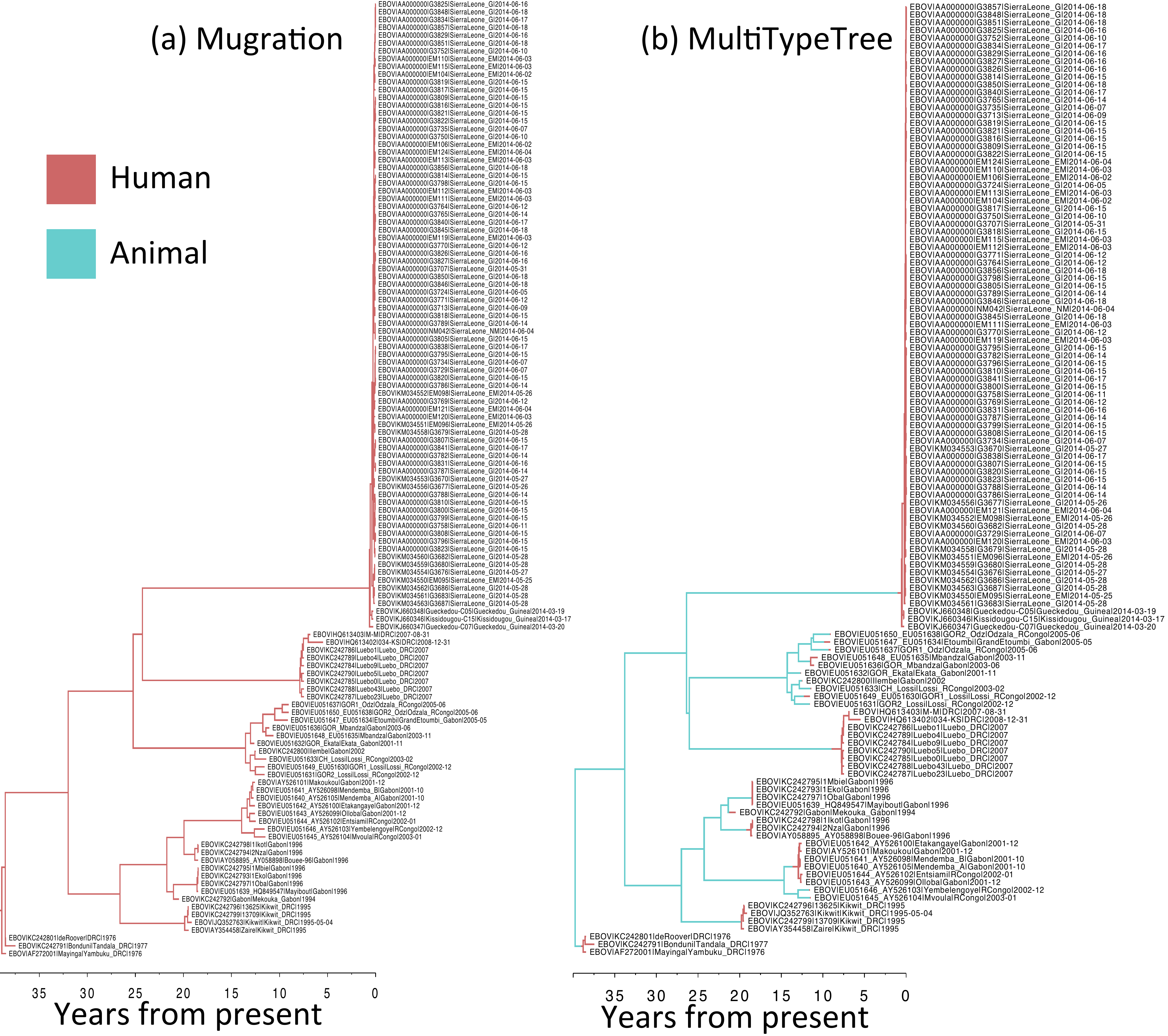}
\label{F6}
\end{figure}

\begin{figure}
\caption{Figure S5}
\includegraphics[width=0.99\textwidth]{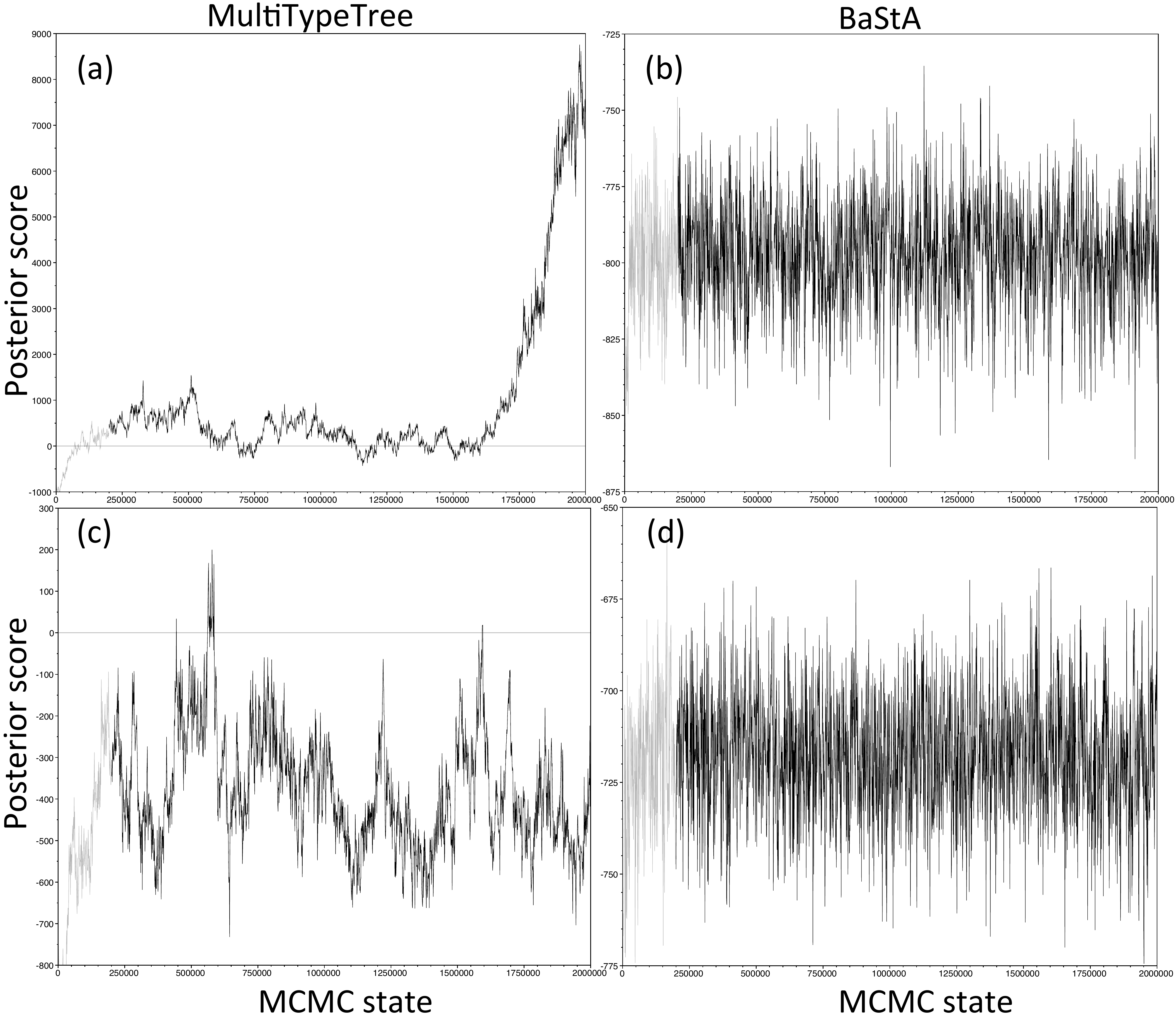}
\label{FS5}
\end{figure}

\end{document}